\documentclass[prb,twocolumn,showpacs,preprintnumbers,amsmath,amssymb]{revtex4}

\usepackage{graphicx}
\usepackage{dcolumn}
\usepackage{bm}

\begin{document}
\addtolength{\topmargin}{2cm}

\title{Manipulating $4f$ quadrupolar pair interactions in TbB$_2$C$_2$ using a magnetic field}

\author{A.M. Mulders$^1$, U. Staub$^1$, V. Scagnoli$^1$,
Y. Tanaka$^2$, A. Kikkawa$^2$, K. Katsumata$^2$ and J.M. Tonnerre$^3$}
 \affiliation{$^1$Swiss Light Source, Paul Scherrer Institut, 5232 Villigen PSI,
Switzerland}
\affiliation{$^2$RIKEN SPring-8 Center, Harima Institute, Sayo, Hyogo 679-5148, Japan}
\affiliation{$^3$CNRS Grenoble, 38042 Grenoble Cedex 9, France}

\date{\today}

\begin{abstract}
Resonant soft x-ray Bragg diffraction at the Tb $M_{4,5}$ edges and non resonant Bragg diffraction
have been used to investigate orbitals in TbB$_2$C$_2$.
The Tb $4f$ quadrupole moments are ordered in zero field below $T_N$ and show a ferroquadrupolar alignment dictated by the antiferromagnetic order.  
With increasing applied field along [110] the Tb $4f$ 
magnetic dipole moments rotate in a gradual manner toward the field. The quadrupole moment is rigidly coupled to the magnetic moment and follows this field-induced rotation.
The quadrupolar pair interaction is found to depend on the specific orientation of the orbitals as 
predicted theoretically and can be manipulated with an applied magnetic field. 
\end{abstract}


\maketitle

Correlation between conduction electrons and electronic orbitals leads to interesting materials properties such as
metal-insulator transitions, colossal magneto resistance and superconductivity. 
Aspheric electronic orbitals, characterized by their quadrupole moment,
may order and cause partial charge localization of the conduction electrons \cite{grenier_prb_2004}
or mediate coupling between cooper pairs.\cite{takimoto_prb_2000}
Orbital order in $f$ electron materials is dominated by coupling with the lattice (Jahn-Teller) or by indirect Coulomb interactions via the conduction electrons. That the latter can be important for intermetallic compounds was established theoretically \cite{teitelbaum_prb_1976, levy_prl_1979}
but detailed experimental knowledge is limited due to the difficulty observing the associated orbital excitations. 
A recent neutron diffraction study provided evidence of a modulated quadrupolar motif in PrPb$_3$, believed to be a direct consequence of indirect Coulomb interactions exhibiting oscillatory nature. \cite{onimaru_prl_2005}
The large orbital momentum of the $f$ electronic shell also gives rise to a significant influence of higher multipole moments and may lead to hidden order phase transitions as demonstrated in 
the extensively studied URu$_2$Si$_2$. \cite{kiss_prb_2005}
Therefore it is important to understand the quadrupolar and higher order multipole pair interactions in these materials.

Quadrupolar order has been successfully investigated using neutron scattering in applied fields where induced magnetic moments reveal the underlying quadrupolar arrangement or motif. In addition, the relatively weak x-ray diffraction intensity of the quadrupole moments can be observed using synchrotron techniques allowing a {\it direct} determination of orbital motifs.
Resonant x-ray scattering at the $L_{2,3}$ edge provided the first proof of the orbital motif in DyB$_2$C$_2$.\cite{tanaka_jpcm_1999, hirota_prl_2000} 
TbB$_2$C$_2$ is proposed
to exhibit a transition from antiferromagnetic (AFM) to antiferroquadrupolar (AFQ) order in an applied magnetic field \cite{kaneko_prb_2003}
and is therefore an interesting candidate for investigation of its orbital interactions.

A magnetic field is time-odd and cannot couple to quadrupole moment which is time-even, but nevertheless in TbB$_2$C$_2$
the ordering temperature increases to 35~K, well 
above $T_N$~=~21.7~K, in an applied magnetic field of 10~T.
Consequently the interplay between dipole (time-odd) and quadrupole (time-even) moments can be readily investigated.
Neutron diffraction studies revealed similarities between the magnetic structure in applied 
fields and the magnetic structure in the combined AFQ and AFM phase of 
DyB$_2$C$_2$.\cite{kaneko_apa_2002} However, resonant x-ray diffraction at the Tb $L_3$ edge suggests AFQ
is present below $T_N$ in zero field with a $k$ vector component of (0,0,$\frac{1}{2}$) as found for DyB$_2$C$_2$.\cite{okuyama_jpsj_2005} 

To test the suggested field-induced quadrupolar ordering we have investigated TbB$_2$C$_2$ 
in an applied magnetic field, $H$, with x-ray scattering techniques. We show that orbital order is present in 
the AFM phase of TbB$_2$C$_2$ at zero field,
however, with ferroquadrupolar alignment which is dictated by the AFM structure.
The quadrupoles are rigidly linked to the magnetic dipole moments and
rotate with $H$ along the [110] direction. We show that, 
due to this rotation, the quadrupole pair interactions become stronger with applied field. 

A single crystal of TbB$_2$C$_2$ was grown by Czochralski
method using an arc-furnace with four electrodes. Samples were cut and polished perpendicular to the [001] direction. 
The (00$\frac{1}{2}$) reflection was recorded at 
the Tb $M_{4,5}$ edges of TbB$_2$C$_2$ at the RESOXS end-station of the SIM beamline at the
Swiss Light Source of the Paul Scherrer Institut.
A permanent magnet provided a field of 1~T parallel 
to the [110] direction in the scattering plane.
In addition, non resonant x-ray Bragg diffraction of the (01$\frac{l}{2}$) and (11$\frac{l}{2}$) reflections was performed at 
the BL19LXU beam line at SPring-8 with 30 keV x-rays using a Ge solid-state detector to
eliminate higher order harmonics. A cryomagnet provided $H$ along the [110] direction.
In both sets of measurements, structure factors were derived from the integrated intensity, corrected for polarization, Lorenz factor,  
absorption and sample geometry. 

\begin{figure}[t]
\includegraphics[width=.7\columnwidth,angle=270]{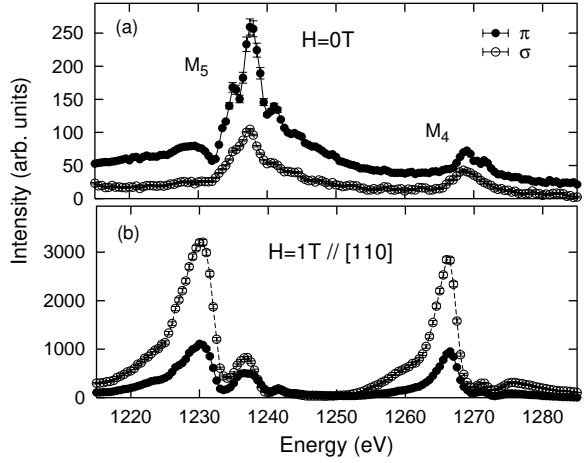}
\vspace{-.3cm}
\caption{Energy profile of the (00$\frac{1}{2}$) reflection at the Tb $M_{4,5}$ edges in TbB$_2$C$_2$ in 
(a) zero field and (b) applied field of 1~T along [110], recorded with $\sigma$ or $\pi$ incident radiation at 11~K.
}
\label{fig_resoxs}
\vspace{-.3cm}
\end{figure}

The resonant Bragg diffraction results, 
Figure \ref{fig_resoxs}a, show the energy dependence of the (00$\frac{1}{2}$) reflection recorded at the Tb $M_{4,5}$ edges in zero field. 
The energy profile is independent of sample rotation about the Bragg wave vector.
A significant part of the diffracted intensity is due to the total reflectivity of the polished sample.
The (00$\frac{1}{2}$) diffracted intensity is 
relatively weak and appears magnetic in origin as witnessed by the relatively large intensity for incoming x-rays polarized 
in the scattering plane ($\pi$). 
In contrast, the diffracted intensity with $H$~=~1~T, along the [110] direction, shown in Fig.~\ref{fig_resoxs}b is much stronger 
and strongest for incoming radiation polarized perpendicular 
to the scattering plane ($\sigma$). The energy profiles shown in Fig.~\ref{fig_resoxs} are expected to be insensitive to the 
magnetic structure,
not to be confused with the total intensity that does depend on the magnetic structure.
The drastic change with $H$ indicates the presence of either an additional component of resonant scattering or a change in relative contribution of two different
components of resonant scattering.
The energy profile recorded in an applied magnetic field is 
similar to that recorded for the (00$\frac{1}{2}$) reflection in DyB$_2$C$_2$ in the AFQ phase.\cite{mulders_tbp} 
This implies that for TbB$_2$C$_2$, under the influence of an applied magnetic field of 1~T,
the (00$\frac{1}{2}$) reflection is dominated by quadrupolar scattering.

\begin{figure}[t]
\includegraphics[width=.8\columnwidth,angle=270]{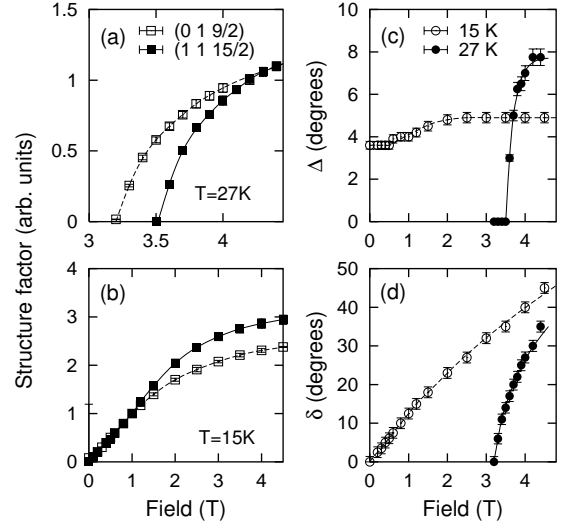}
\vspace{-.3cm}
\caption{Structure factor of the (01$\frac{9}{2}$)
and (11$\frac{15}{2}$) reflection as function of applied magnetic field recorded with 30 keV at (a) T~=~27~K and (b) T~=~15~K.
(c) and (d) show the deduced angles of Tb $4f$ orbital rotation as defined in Fig.~\ref{fig_angles}.
}
\label{fig_SP8}
\vspace{-.3cm}
\end{figure}

The non resonant Bragg diffraction results illustrate that the quadrupole moment follows the magnetic moment rotation in applied fields.
Fig.~\ref{fig_SP8}a,b shows the structure factors that relate directly to the order parameter, obtained from the integrated, non-resonant Bragg intensities of the (01$\frac{9}{2}$) and (11$\frac{15}{2}$) reflection as a function of applied field.
In the AFM phase, at 15~K, the structure factors show a linear increase for applied fields below 1.5~T.
In the paramagnetic phase, at 27~K, the ordered phase is entered above 3~T. The (01$\frac{9}{2}$) reflection appears at 
3.2~T while the (11$\frac{15}{2}$) reflection appears at 3.5~T. However the evolution of the structure factor with increasing field is 
smooth for both reflections without a hint of a phase 
transition at 3.5~T for (01$\frac{9}{2}$).
A similar behavior is observed as function of temperature in an applied field of 5~T (not shown). 
The intensities of (01$\frac{9}{2}$) and (11$\frac{15}{2}$) show a smooth decrease with increasing temperature and disappear
below 30~K and 29.5~K, respectively.

A structural transition with alternating displacements of the B and C atoms gives intensity in both type of reflections\cite{adachi_prl_2002} and cannot account for the present observation of intensity at (01$\frac{9}{2}$) and the absence of intensity at (11$\frac{15}{2}$).
Magnetic, or time-odd, x-ray diffraction is generally weak and does not explain the magnitude of the observed intensity. 
We conclude that the scattering arises from time-even x-ray diffraction, i.e. from the aspheric charge distribution of the $4f$ shell, as such scattering can be intense. 

\begin{figure*}[t]
\includegraphics[width=.67\columnwidth,angle=270]{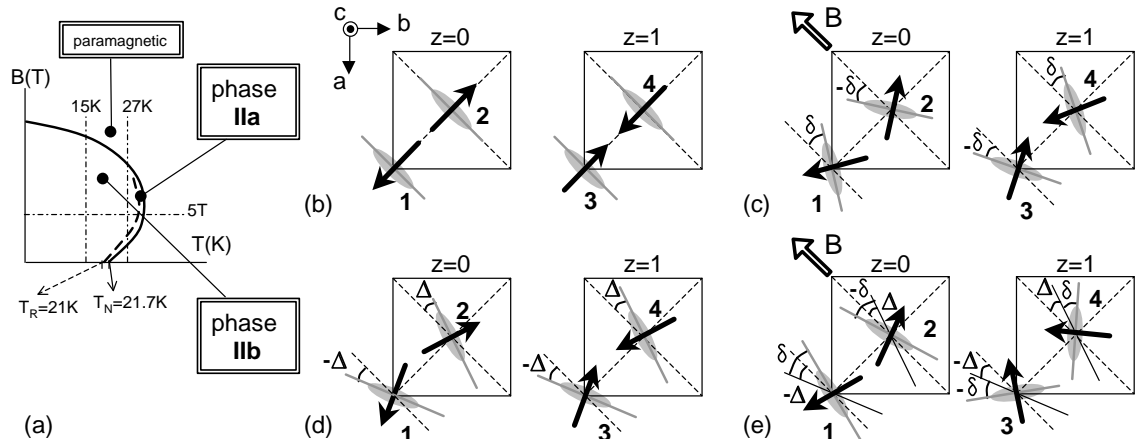}
\vspace{-.3cm}
\caption{(a) Phase diagram of TbB$_2$C$_2$ with magnetic field applied along [110]. (b) and (c) illustrate the orientation of magnetic 
moments (arrows) and 
quadrupole moments (ovals) in TbB$_2$C$_2$ in phase IIa in zero field and under applied fields, respectively. (d) and (e) illustrate phase 
IIb under the same conditions.
}
\label{fig_angles}
\vspace{-.3cm}
\end{figure*}

To determine the structure factor it is assumed that the Tb ion on each site has the same charge distribution except for its specific orientation. 
Using the formalism for non resonant time-even scattering of Lovesey {\it et. al.}\cite{lovesey_review} we deduce two 
independent Tb ions contribute and obtain for the structure factors:
\begin{eqnarray}
F_{1 1 \frac{l}{2}}\propto -2 \left[ \sin (2\phi_1) -  \sin (2\phi_3) \right] \langle  Q_{1 1 \frac{l}{2}} \rangle \label{eq1},\\
F_{0 1 \frac{l}{2}} \propto   \left[ \cos (2\phi_1) - \cos (2\phi_3) \right] \langle Q_{0 1 \frac{l}{2}} \rangle,
\label{eq2}
\end{eqnarray}
\noindent 
when the Miller index $l$ equals an odd integer. The principal axis of the $4f$ orbital at atom $n$ is canted in the $ab$-plane with respect 
to the [100] direction with angle $\phi_n$.
$\langle  Q_{h k l}  \rangle$ is the expectation value of the Tb $4f$ time-even multipolar moment for a given field, temperature and
Bragg wave vector {\bf k} = ($k_a$,$k_b$,$k_c$):
\begin{eqnarray}
\langle Q_{1 1 \frac{l}{2}} \rangle \propto \frac{k_b^2}{k^2} \left[ \langle j_2 \rangle \Psi_2^2 + 3\sqrt{3} \langle j_4 \rangle  \Psi_2^4  
 - \sqrt{182} \langle j_6 \rangle \Psi_2^6  \right],\label{eq3}\\
\langle Q_{0 1 \frac{l}{2}} \rangle \propto \langle Q_{1 1 \frac{l}{2}} \rangle+ \frac{k_b^2}{k^2} \sqrt{21} \langle j_4 \rangle  \Psi_4^4.
\label{eq4}
\end{eqnarray}
where $\Psi_q^x$ is a structure factor of the chemical unit cell that is a linear sum of the atomic tensors, 
$\Psi_q^x$=$\langle T_q^x \rangle - \langle T_{-q}^x \rangle$. $\langle T_q^x \rangle$ describes the asphericity of the 
Tb $4f$ electronic shell where
$\langle T_2^2 \rangle$ equals the quadrupole moment,  $\langle T_2^4 \rangle$ the hexadecapole moment 
and $\langle T_2^6 \rangle$ the hexacontatetrapole moment. 
$\langle j_2 \rangle$, $\langle j_4 \rangle$ and $\langle j_6 \rangle$ are Bessel function transforms of the radial distribution
of the $4f$ electronic charge.
In eqs.~\ref{eq3} and \ref{eq4}, ($k_c$/$k$)$^2\sim1$ and we neglect higher order terms ($k_b$/$k$)$^4$
and ($k_b$/$k$)$^6$ as $k_b$/$k\ll$1. 

We discuss the configuration of the quadrupoles but the order of the higher multipoles may alter also.
When the $4f$ quadrupoles are aligned along [110] ($\phi_1$=$\phi_3$=$\frac{3}{4}\pi$) as drawn in Fig.~\ref{fig_angles}(b), 
$F_{01\frac{l}{2}}$ and $F_{11\frac{l}{2}}$ are both zero. 
When the $4f$ quadrupoles tilt toward the applied field with angle 
$\delta$ ($\phi_1$=$\frac{3}{4}\pi + \delta$; $\phi_3$=$\frac{3}{4}\pi - \delta$), Fig.~\ref{fig_angles}(c), 
$F_{01\frac{l}{2}}$ is finite and $F_{11\frac{l}{2}}$ remains zero.
Neutron diffraction shows that in the AFM phase the dipole moments tilt away 
from $\langle$110$\rangle$ with angle $\Delta$, Fig.~\ref{fig_angles}(d) 
($\phi_1$=$\frac{3}{4}\pi - \Delta$; $\phi_3 $=$ \frac{3}{4}\pi - \Delta$).\cite{kaneko_jpsj_2001}  
In this case, both $F_{01\frac{l}{2}}$ and $F_{11\frac{l}{2}}$ are zero in zero field while under an 
applied field both become finite (Fig.~\ref{fig_angles}(e), 
$\phi_1$=$\frac{3}{4}\pi - \Delta + \delta$; $\phi_3$=$\frac{3}{4}\pi - \Delta - \delta$). 

Consequently, at 27~K, between $H$~=~3.2 and $H$~=~3.5~T, the magnetic and
quadrupolar structure of TbB$_2$C$_2$ is characterized by phase IIa. The quadrupoles are aligned 
parallel to [110] as illustrated in Fig.~\ref{fig_angles}(b) and are tilted in applied field as
illustrated in Fig.~\ref{fig_angles}(c). Above 3.5~T the compound is characterized by phase IIb. The quadrupoles are tilted away 
from the [110] direction and neighboring atoms in the $ab$-plane are tilted in opposite directions as illustrated for $H$~=~0 in
Fig.~\ref{fig_angles}(d) and for $H$~$\neq$~0 in Fig.~\ref{fig_angles}(e).
We show the phase diagram schematically in Fig.~\ref{fig_angles}(a) and define a reorientation temperature $T_R$ by the boundary between
phases IIa and IIb.

Multipole moment rotation in an applied field is linear for small $\delta$, $\Delta$ (see Eq.~(\ref{eq1})) hence the
linear increase of the structure factor as observed at 15~K (see Fig.~\ref{fig_SP8}(b)). 
The magnetic moment of the $4f$ shell tilts toward the direction of the magnetic field and the time-even multipole moments follow accordingly, giving the observed contrast. This shows that (i) the coupling between magnetic and quadrupole moment
is rigid and (ii) the orbitals are already ordered in zero field, although with a ferroquadrupolar alignment.
This is consistent with the result in Fig.~\ref{fig_resoxs}a as the ferroquadrupolar 
alignment in zero field does not contribute to scattering at (00$\frac{1}{2}$), whereas in an applied field, due to the orbital rotation, 
time-even scattering dominates (Fig.~\ref{fig_resoxs}b).

\begin{figure}[t]
\includegraphics[width=.3\columnwidth,angle=270]{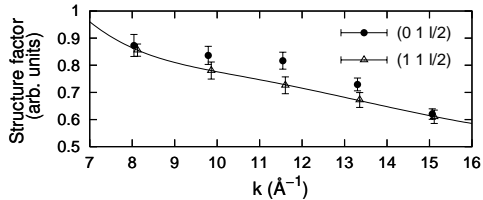}
\vspace{-.3cm}
\caption{Structure factor of (01$\frac{l}{2}$) and (11$\frac{l}{2}$) reflections at B=5~T and T=2~K.
The solid line is a fit to Eq.~\ref{eq1}.
}
\label{fig_q}
\vspace{-.3cm}
\end{figure}

The normalized structure factor of (01$\frac{l}{2}$) and (11$\frac{l}{2}$) reflections at 5~T
with $9 \leq l \leq 17$ and are presented in Fig~\ref{fig_q}. 
They are equal within the experimental uncertainty which shows that $\Psi_4^4$ is small (Eqs.~\ref{eq3} and \ref{eq4}). 
The solid line corresponds to Eqs.~\ref{eq1} and \ref{eq3} with $\Psi_2^4/\Psi_2^2$=0.5 and $\Psi_2^6/\Psi_2^2$=0.5 and shows that the 
intensity can be attributed to time-even x-ray diffraction. The limited range of $k$ makes the determination of 
$\Psi_2^2$ uncertain and this analysis merely confirms the absence of scattering from B and C atoms in contrast to 
DyB$_2$C$_2$.\cite{adachi_prl_2002} Thus orbital order is present independent of a structural transition in TbB$_2$C$_2$ verifying the orbital interaction is mediated via the conduction electrons. A similar conclusion was reached for DyB$_2$C$_2$ from inelastic neutron scattering.\cite{staub_prl_2005}

With $\langle Q_{01\frac{l}{2}} \rangle  = \langle Q_{11\frac{l}{2}} \rangle $ for a given field and temperature
the angle $\Delta$ is deduced from the ratio between $F_{01\frac{9}{2}}$ and $F_{11\frac{15}{2}}$.
In addition the quantity $\sin (2\delta) \langle Q_{hkl} \rangle$ follows from Eqs.~\ref{eq1} and \ref{eq2} and $\delta$ has 
been estimated assuming $\langle Q_{hkl} \rangle$ is constant as a function of field and temperature.
Both results are presented in Figs.~\ref{fig_SP8}(c) and (d). 
If $\langle Q_{hkl} \rangle$ increases with $H$ the actual values for $\delta$ are lower than presented in Fig.~\ref{fig_SP8}d.
Nevertheless, the increase in $\delta$ is gradual as a function of the applied field and reflects the orbital moment
rotation. 

The angle $\Delta$ is a result of the competition between the indirect Coulomb interaction and the magnetic exchange interaction between neighboring ions in the $ab$-plane.  The first interaction favors
perpendicular alignment of the quadrupoles while the latter favors a parallel alignment.
An increase in $\Delta$ is observed with $H$. This shows that the relative strength of the indirect Coulomb interaction {\it increases} as the quadrupoles rotate perpendicular to each other in an applied field.
To the authors knowledge, this is the first time that an increase in indirect Coulomb interaction has been observed in a straightforward manner. We also note that the determination of $\Delta$ is robust against changes in $\langle Q_{hkl} \rangle$ with applied field.

The increase in ordering temperature with applied field\cite{kaneko_prb_2003} is in line with an antiferroquadrupolar interaction which becomes stronger when the angle between the orbitals moves from parallel to perpendicular alignment. 
The magnetic moments interact via the polarized spins of the conduction electrons while the multipolar moments
interact via the charge density of the conduction electrons. In other words, charge screening by the conduction electrons changes when 
the $4f$ orbitals rotate and consequently the magnitude of the quadrupole pair interactions changes. 
This is consistent with the theoretical framework of Teitelbaum and Levy \cite{teitelbaum_prb_1976} where electric multipole coupling via the conduction electrons depends on the relative angular positions of the ion cores.

The occurrence of combined AFQ and AFM below $T_N$ is consistent with the quasi doublet ground-state.\cite{haino_jpsj_1838} 
Removal of the degeneracy of a doublet ground-state may account for one phase transition.
Relaxation between the two singlets takes place above $T_N$ and the magnetic and quadrupolar moments fluctuate.
The increased energy separation between the two singlets below $T_N$ is due to the magnetic exchange interaction, but both the 
magnetic and quadrupolar moment of the ground state singlet are observed, and this is illustrated by this study. 

In conclusion, no magnetic field induced ordering of quadrupoles exists in TbB$_2$C$_2$. The orbitals are already ordered in the AFM phase 
in zero field with a ferroquadrupolar motif.
On increasing the applied field along [110] the Tb $4f$ magnetic moments rotate toward the field direction in a gradual manner. 
The quadrupole moment is rigidly coupled to the magnetic moment and follows
this rotation. Neighboring Tb $4f$ orbitals move from parallel to perpendicular alignment and the quadrupolar pair interaction
increases, witnessed by an increase of angle $\Delta$ between neighboring quadrupoles. 
This study shows that the quadrupolar pair interaction depends on the specific orientation of the orbitals as predicted for indirect Coulomb interaction via the conduction electrons and can be 
manipulated with an applied field.

We thank S.W. Lovesey for valuable discussion. This work was supported by the Swiss National Science Foundation and 
by a Grant-in-Aid for Scientific Research from the Japan Society for the Promotion of 
Science. This work was in part performed at the SLS of the Paul Scherrer Institute, Villigen PSI, Switzerland.

\end{document}